\documentclass[fleqn,usenatbib]{rasti}

\usepackage{newtxtext,newtxmath}

\usepackage[T1]{fontenc}

\DeclareRobustCommand{\VAN}[3]{#2}
\let\VANthebibliography\thebibliography
\def\thebibliography{\DeclareRobustCommand{\VAN}[3]{##3}\VANthebibliography}

\usepackage{bm}
\usepackage{multirow}
\usepackage{wrapfig}
\usepackage{graphicx}	% Including figure files
\usepackage{amsmath}	% Advanced maths commands
\title[Bayesian inference in Astronomy]{Practical Guidance for Bayesian Inference in Astronomy}

\author[G.M. Eadie et al.]{Gwendolyn M.~Eadie$^{1,2}$\thanks{E-mail: gwen.eadie@utoronto.ca},
Joshua S. Speagle,$^{1,2,3}$
Jessi Cisewski-Kehe,$^{4}$
Daniel Foreman-Mackey,$^{5}$ \newauthor
Daniela Huppenkothen,$^{6}$
David E. Jones,$^{7}$
Aaron Springford$^{8}$
and Hyungsuk Tak$^{9, 10,11}$
\\
$^{1}$University of Toronto, David A. Dunlap Department of Astronomy \& Astrophysics, Toronto, M5S 3H4, Canada\\
$^{2}$University of Toronto, Department of Statistical Sciences, Toronto, M5S 3G3, Canada \\
$^{3}$University of Toronto, Dunlap Institute for Astronomy \& Astrophysics, Toronto, M5S 3H4, Canada\\
$^{4}$University of Wisconsin-Madison, Department of Statistics, Madison, WI, 53706, USA\\
$^{5}$Center for Computational Astrophysics, Flatiron Institute, 160 5th Ave, New York, NY 10010, USA\\
$^{6}$SRON Netherlands Institute for Space Research, Niels Bohrlaan 4, 2333 CA Leiden, Netherlands\\
$^{7}$Texas A\&M University, Department of Statistics, College Station, TX 77843, USA\\
$^{8}$Cytel, Toronto, Ontario, Canada\\
$^{9}$Pennsylvania State University, Department of Statistics,  University Park, PA 16802, USA\\
$^{10}$Pennsylvania State University, Department of Astronomy \& Astrophysics,  University Park, PA 16802, USA\\
$^{11}$Pennsylvania State University, Institute for Computational and Data Sciences,  University Park, PA 16802, USA
}

\begin{document}
\label{firstpage}
\pagerange{\pageref{firstpage}--\pageref{lastpage}}
\maketitle

\begin{abstract}
In the last two decades, Bayesian inference has become commonplace in astronomy. At the same time, the choice of algorithms, terminology, notation, and interpretation of Bayesian inference varies from one sub-field of astronomy to the next, which can lead to confusion to both those learning and those familiar with Bayesian statistics. Moreover, the choice  varies between the astronomy and statistics literature, too. In this paper, our goal is two-fold: (1) provide a reference that consolidates and clarifies terminology and notation across disciplines, and (2) outline practical guidance for Bayesian inference in astronomy. Highlighting both the astronomy and statistics literature, we cover topics such as notation, specification of the likelihood and prior distributions, inference using the posterior distribution, and posterior predictive checking. It is not our intention to introduce the entire field of Bayesian data analysis -- rather, we present a series of useful practices for astronomers who already have an understanding of the Bayesian "nuts and bolts" and wish to increase their expertise and extend their knowledge. Moreover, as the field of astrostatistics and astroinformatics continues to grow, we hope this paper will serve as both a helpful reference and as a jumping off point for deeper dives into the statistics and astrostatistics literature.
\end{abstract}


%%%%%%%%%%%%%%%%% BODY OF PAPER %%%%%%%%%%%%%%%%%%

\section{Introduction}
Over the past two decades, Bayesian inference has become increasingly popular in astronomy. On NASA's Astrophysics Data System (ADS), a search using ``keyword:statistical'' and ``abs:bayesian'' yields 2377 refereed papers, and shows exponential growth since the year 2000, with over 237 papers in 2021.

Bayesian analyses have become popular in astronomy due to several key advantages over traditional methods. First, an estimate of the posterior distribution of model parameters provides a more complete picture of parameter uncertainty, joint parameter uncertainty, and parameter relationships given the model, data, and prior assumptions than traditional methods. Second, the interpretation of Bayesian probability intervals is often closer to what scientists desire, and is an appealing alternative to point estimates with confidence intervals which often rely on the sampling distribution of the estimator. Third, Bayesian analysis easily allows for marginalization over nuisance parameters, incorporation of measurement uncertainties through measurement error models, and inclusion of incomplete data such as missing and censored data.  Fourth, astronomers often have prior knowledge about allowable and realistic ranges of parameter values (e.g., through physical theories and previous observations/experiments) which can naturally be included in prior distributions and thereby improve the final inference. 

Importantly, in addition to the aforementioned advantages of the Bayesian approach, efficient and increased computing power, along with easy-to-use or out-of-the-box algorithms, have brought Bayesian methodology to astronomers in convenient practical forms (e.g., \textit{emcee} \citep{Foreman_Mackey_2013}, \textit{Rstan}~\citep{RStan}, \textit{PyStan}~\citep{allen_riddell_2017_1003176}, \textit{PyMC3}~\citep{pymc3},  BUGS \citep{lunn2000winbugs}, NIMBLE~\citep{nimble}, and JAGS~\citep{plummer2003jags}).

Interestingly, the surge in popularity of Bayesian statistics comes in spite of the fact that Bayesian methods are rarely taught in undergraduate astronomy and physics programs, and has only recently been introduced at a basic level in astronomy graduate courses \citep{2019clrpWP, Astro2020WP}. Some challenges faced by both new and seasoned users of Bayesian inference are the varied notation, terminology, interpretation, and choice of algorithms available in the astronomy and statistics literature.

Being well-versed in best practices and common pitfalls associated with the Bayesian framework is important if these methods are to be used to advance the field of astronomy. Here, users of Bayesian inference in astronomy face challenges too, since undergraduate and graduate program training is still catching up to the state-of-the-art Bayesian inference methods.

The goal of this paper is two-fold. Our first goal is to provide a ``translation'' between terminology and notation used for Bayesian inference across the fields of astronomy and statistics. Our second goal is to illustrate useful practices for the Bayesian inference process which we hope will be a valuable contribution to astronomers who are familiar with and/or use Bayesian statistics in research. To achieve these goals, we  deal with the following topics in the main body of the paper: notation (Section~\ref{sec:notation}), interpreting and determining the likelihood (Section~\ref{sec:likelihood}), choosing and assessing prior distributions (Section~\ref{sec:prior}), evaluating and making inference from the posterior distribution (Section~\ref{sec:posterior}), and performing posterior predictive checks (Section~\ref{sec:pcheck}). 

This work is not meant to be a comprehensive introduction to Bayesian inference, but rather an unveiling of Bayesian statistics as both an extensive topic and an active research area.  We focus our efforts on identifying common mistakes and misunderstandings related to Bayesian inference, and use these as jumping off points for highlighting important topics for further study. Indeed, many valuable topics and subtopics arise which we do not cover, but we make a point of providing key references. For example, throughout the paper we touch on areas such as Bayesian design, posterior predictive checking, hierarchical modeling, and Bayesian computations \citep{craiu2014bayesian, robert2014bayesian}, and also  recommend books on Bayesian data analysis from statistics and astrostatistics \citep{gelman2013bayesian, carlin2008bayesian,hilbe2017bayesian}.

To help the narrative, we use a running example at the end of each section --- inferring the distance to a star through its parallax measurement. The specific problem of distance estimation from parallax in the Bayesian context is explored closely in other studies \citep{2015PASP..127..994B,2016ApJ...832..137A,2016ApJ...833..119A} and applied to the \textit{Gaia} second data release (\textit{Gaia} DR2) \citep{Lindegren, BailerJones2018AJ,schonrich2019}, and we refer the reader to these papers for a deep exploration on this topic. Here, we employ this example because of its generality, because it provides some interesting challenges and potential pitfalls, and because it provides a nice framework to illustrate sound practices in Bayesian analysis. Along the way, we also identify how our advice applies to other areas in astronomy.

\section{Specifying a Bayesian Model}\label{sec:setup}

\subsection{Notation \& Bayes Theorem}\label{sec:notation}

A number of different notation practices for Bayesian inference are used in both the astronomy and statistics literature. This section is meant to clarify some of these differences, while also providing a ``translation'' so that astronomers can more easily follow statistics papers (e.g., recognize notation for random variables, probability distribution functions, etc.). 

We use $\bm{y}$ (vectorized form) to represent the observed data of a random variable $Y$, and $\bm{\theta}$ to represent the parameter(s) of interest. The posterior distribution is defined by Bayes' theorem as
\begin{equation}
    p(\bm{\theta} | \bm{y}) = \frac{p(\bm{y}|\bm{\theta})p(\bm{\theta})}{p(\bm{y})}
    \label{eq:Bayes}
\end{equation}
where $p(\bm{y}|\bm{\theta})$ is the \textit{sampling distribution} for $\bm{y}$ given $\bm{\theta}$ (Section \ref{sec:likelihood}), $p(\bm{\theta})$ is the \textit{prior density} (Section \ref{sec:prior}), and $p(\bm{y})$ is the \textit{prior predictive density} \citep{schervish1995theory}. With the data $\bm{y}$ in hand, $p(\bm{y}|\bm{\theta})$ is often
viewed as a function of $\bm{\theta}$ called the \emph{likelihood function} (which is not a probability density), and $p(\bm{y})$
is a normalizing constant that does not depend on $\bm{\theta}$ (which is often referred to in astronomy as the model evidence). In the probability and Bayesian computation literature, the posterior probability distribution of interest is usually denoted and referred to as the \textit{target distribution} with the notation $\pi$. There are also differences in notation for the likelihood across disciplines, which we discuss in Section~\ref{sec:likelihood}.

For smooth translation between sub-disciplines of astronomy and statistics, it is beneficial to use explicit statements about model choices and parameter definitions. For example, a list of all model parameters, notation, and their associated prior probability distributions in the form of a table is very useful to the reader. Moreover, we stress the importance of fully specifying any Bayesian model in papers to increase reproducibility (e.g., via a detailed appendix, open code). In this spirit, we provide the full Bayesian model for our running example in Table~\ref{tab:everything}, explicitly define notation next, and provide our open source code\footnote{\url{https://github.com/joshspeagle/nrp_astrobayes}}. 

\subsubsection{Parallax Example: Defining Notation}

For the running example in this paper, we infer the distance to a star from a parallax measurement. The true but unknown distance $d$ in kiloparsecs (kpc) is related to the true but unknown parallax $\varpi$ in milliarcseconds (mas) through 
\begin{equation}
    d \,[{\rm kpc}] = \frac{1}{\varpi \,[{\rm mas}]}.
    \label{eq:distvar}
\end{equation}
Our data $y$ is a measurement of the parallax, and has some fixed uncertainty $\sigma$ that we treat as known. Thus, in the Bayesian framework, we wish to infer the parameter $d$ given the data $y$, and we seek to find the posterior distribution,
\begin{equation}
p(d|y) \propto p(y|d)p(d). 
\end{equation}
In Section~\ref{sec:extendedexample}, we extend this example to infer the distance to a \textit{star cluster} from the parallax measurements of many stars within the cluster. The true but unknown distance to the cluster, $d_{\rm cluster}$, is related to the true but unknown parallax of the cluster through Equation~\ref{eq:distvar}, too. In this case, we express the corresponding posterior density as follows:
\begin{equation}
p(d_{\rm cluster}|\bm{y}) \propto p(\bm{y}|d_{\rm cluster})p(d_{\rm cluster}),
\end{equation}
where $\bm{y}$ represents a vector of parallax measurements of its stars. Table~\ref{tab:everything} summarizes this model specification.
%to denote the posterior distribution, 

\begin{table*}
\caption{Bayesian models for inferring (1) a star's distance parameter $d$ from its parallax measurement $y$, assuming a Gaussian distribution for its true parallax $\varpi$ (top panel), and (2) a star cluster's distance parameter $d_{\textrm{cluster}}$ from the parallax measurement of  $n$ stars $\bm{y}$ (bottom panel).\label{tab:everything} }
\centering
\begin{tabular}{|l|c|}
\hline
\textbf{Inferring a single star's distance parameter $d$} & \\
\hline 
  Sampling density / likelihood (parallax) & $p(y \mid d) = N(y\mid 1/d, \sigma^2)$, where $\varpi = 1/d$ \\
  &\\
  Prior (distance) & $p(d) \propto 
    \begin{cases}
    d^2 e^{-d/L} & \textrm{if~} d_{\rm min} < d < d_{\rm max} \textrm{with a constant $L$}\\
    0 & {\rm otherwise} 
    \end{cases}$\\
  & \\
  Posterior on distance & $p(d | y, \sigma_\varpi) \propto d^2 \exp\left[{-\frac{d}{L}-\frac{(y - 1/d)^2}{2\sigma^2}}\right]$ \\
  \hline
  \hline
  \textbf{Inferring a star cluster's distance parameter $d_{\textrm{cluster}}$} & \\
  \hline
  Sampling density / likelihood ($n$ parallax measurements) &  $p(\bm{y} \mid d_{\textrm{cluster}}) = \prod^{n}_{i=1} N(\bm{y}\mid 1/d_{\textrm{cluster}}, \sigma^2_i) $ \\
  \hline
  Prior (on parallax of cluster) & $p(\varpi_{\textrm{cluster}})$ = $N(\varpi_{\textrm{cluster}}\mid \mu_{\varpi}, \sigma_{\varpi})$\\
  \hline
  Posterior on distance to cluster & $ p(d_{\rm cluster} | \bm{y}) \propto p(d_{\rm cluster})  \exp\left[{-\frac{(y_{eff} - 1/d_{\textrm{cluster}})^2}{2\tau^2}}\right]$ \\
  \hline
\end{tabular}

\footnotemark{\small In both the upper and lower box, $\sigma$ and $\sigma_i: i=1,2,\ldots,n$ are assumed to be known. Throughout $N(y\mid a, b)$ denotes the Gaussian density function of $y$ with mean $a$ and variance $b$. \normalsize} 
\end{table*}

\subsection{Likelihood function}\label{sec:likelihood}

 Differences in notation and language between statistics and astronomy can lead to confusions regarding the likelihood. In Bayesian statistics, the capital letters $Y$ and $\Theta$ often denote \textit{random variables}. For example, both  \cite{gelman2013bayesian} in their applied statistics text and  \cite{schervish1995theory} in his statistics theory text first write down a \textit{joint} probability density $p(\Theta,Y)$, and  then specify the likelihood function as $p(Y=\bm{y}\mid \bm{\theta})$ (sometimes written $\mathcal{L}(\bm{\theta})$ elsewhere), where $\bm{y}$ is the fixed, observed value of $Y$ (i.e., the data), and $\bm{\theta}$ is the argument of the likelihood function. That is, the likelihood is a function of the parameters $\bm{\theta}$, given the data $\bm{y}$ \citep{gelman2013bayesian,schervish1995theory,carlin2008bayesian,berger1988likelihood,casella2002statistical}. However, in astronomy, it is not unusual to see phrases such as ``the likelihood function of the data given the model parameters'', which might be misconstrued as treating the likelihood function as a function of data. Finally, we note that  the likelihood function of $\bm{\theta}$ is not a probability density function of $\bm{\theta}$. %in many cases.  It is a probability density only if 
% \begin{equation}
%  \int p(\bm{y}\mid \bm{\theta}) d\bm{\theta}<\infty,~~\textrm{or equivalently}~\int \mathcal{L}(\bm{\theta}) d\bm{\theta}<\infty.
% \end{equation}
 
 %, it is safe to treat it as a function of
 %, which is a common misconception
 %In short,
%  We  note that the likelihood function is not 
%the sampling density of $\bm{y}$
%be interpreted and described  as 
%Describing the likelihood as both ``given the parameters'' and ``as a function of the parameters'' can lead to confusion. 

\begin{table*}[t!]
    \centering
    \caption{Likelihood notation found in different contexts.}
    \begin{tabular}{|c|l|l|}
    \hline
    \textbf{Notation }& \textbf{Description} & \textbf{Context} \\
    \hline
       $f_Y(\bm{y}|\bm{\theta})$  & distribution function of the random variable $Y$, & statistics  \\
       &  but viewed as a function of $\bm{\theta}$, with $\bm{y}$ fixed &  \\
       \hline
       $p(\bm{y}|\bm{\theta})$  & format used in this paper (common) & statistics and astronomy \\
       \hline
       $\mathcal{L}(\bm{\theta};\bm{y})$ or $\mathcal{L}(\bm{\theta})$ & explicit notation for the likelihood with  
       & specific statistics topics, e.g.,  \\
       & argument $\bm{\theta}$ & maximum likelihood estimation \\
       \hline
       $P(\bm{D}|\bm{\theta})$ & $D$ represents the data & astronomy \\
       \hline
       $P(\bm{y}|M)$ & $M$ represents the model assumption, & astronomy, model selection\\
       & implicitly suggests parameters & \\
       \hline
       $Pr(\bm{y}|\bm{\theta}, H)$ & $H$ represents a particular proposed model & astronomy, model selection \\
       \hline
    \end{tabular}
    \label{tab:likelihood}
\end{table*}

In Table~\ref{tab:likelihood}, we summarize common notation for the likelihood found in both the statistics and astronomy literature, which ranges from being very explicit (e.g., $f_Y(\bm{y}|\bm{\theta})$) to quite simplified (e.g., $P(\bm{y}|M)$). We note that a subtlety sometimes missed in astronomy is the difference between $p$ and $P$. In statistics, $f$ and $p$ are often used for probability density functions (pdf) of continuous random variables, and $P$ or $Pr$ are used to denote probabilities of discrete events (probability mass functions (pmf) are an exception, and are often denoted using $f$, $p$, or $P$). A capital $F$ is usually reserved for the cumulative distribution function (cdf).

Determining what the likelihood function should be in  a given astronomy problem can be challenging, and care must be taken to choose an appropriate sampling distribution. The likelihood is often taken to be a product of independent and identically distributed (i.i.d.) Gaussian random variables with known variance. While this choice is sometimes plausible, there are also many cases in which it is inappropriate, and has a material effect on inference. For instance, when describing the brightness of a high energy source, a discrete distribution such as the Poisson distribution is usually more appropriate than a Gaussian. In other cases, uncertainty in the variance of the data might lead us to use a likelihood function based on the $t$-distribution or another non-Gaussian parametric family. A further consideration is whether the data being modeled are collected as a function of space or time, in which case the assumption of exchangeability -- \textit{that data can be reordered without affecting the likelihood} -- is generally unwarranted. In these cases, an expanded model that includes correlation among observations should be considered.

Best practice includes all non-negligible contributors to the measurement process in the likelihood function. For example, it is important to account for substantial truncation and censoring issues when present, because these can strongly influence parameter inference in some cases \citep{rubin1976inference,EadieWebbRosenthal2021}. Other common issues to check for and address are measurement uncertainty, correlated errors, measurement bias, sampling bias, and missing data. There are a number of valuable references in the statistics literature on these topics \citep{rubin1976inference,little2019statistical}. We recommend writing down enough mathematical details to uniquely determine the likelihood by defining (algebraically) not only the physical process of interest but also the sampling/measurement process that generated the data.

\subsubsection{Parallax Example: the likelihood function}\label{sec:parallax}

The \textit{Gaia} spacecraft has measured parallaxes for over a billion stars \citep{gaia2018gaia, gaia2018vizier}. These parallaxes have been shown empirically, through simulations \citep{2012A&A...543A..15H,2012A&A...538A..78L}, to follow a normal distribution with mean equal to the true underlying parallax $\varpi$ so that
\begin{align}
     p(y \mid \varpi) &= \frac{1}{\sqrt{2\pi\sigma^2}} \exp\left[{-\frac{(y - \varpi)^2}{2\sigma^2}}\right] \textrm{  or equivalently,} \\
     y\mid  \varpi, \sigma &\sim N(\varpi,~\sigma^2)
     \label{eq:model}
\end{align}
where $y$ is the measured parallax and $\sigma$ is the associated (assumed known) measurement uncertainty \citep{hogg2018likelihood}. The parameter of interest is the distance $d$, so we rewrite Equation~\ref{eq:model} as
\begin{align}
    p(y \mid d) &= \frac{1}{\sqrt{2\pi\sigma^2}} \exp\left[{-\frac{(y - 1/d)^2}{2\sigma^2}}\right] \textrm{  or equivalently,} \\
    ~~y\mid  d, \sigma &\sim N(1/d,~\sigma^2).
    \label{eq:likelihood}
\end{align}

We note that  a similar Gaussian model assumption is widely applicable to various sub-fields in observational astronomy such as detecting exoplanets by RV \citep{danby1988fundamentals,mayor2011harps,pepe2011harps,fischer2013twenty,butler2017lces} or by transit \citep{konacki2003extrasolar,alonso2004tres,dragomir2019tess}, inferring the true brightness of a source \citep{tak2017bayesian}, or estimating the Hubble constant  \citep{1929PNAS...15..168H}. This is because the statistical details are analogous; the observation is measured with Gaussian measurement error,  estimated measurement error uncertainty $\sigma$ is treated as a known constant, and the mean model can be written as a deterministic function of other parameters, e.g., $\varpi=1/d$ in Equation~\ref{eq:model}. On the other hand, as  mentioned already, in each new setting it is important to carefully consider which model is most appropriate; Gaussian-based models are (i)  sometimes misused, (ii) overused, or (iii) sometimes inappropriate.

\subsection{Prior Distributions}\label{sec:prior}

The prior probability distribution, or \emph{the prior}, captures our initial knowledge about the model parameters before we have seen the data. Priors may assign higher probability (or density) to some values of the model parameters over others. Priors are often categorized into two classes: \textit{informative} and   \textit{non-informative}. The former type summarizes knowledge gained from previous studies, theoretical  predictions, and/or scientific intuition.  The latter type attempts to include as little information as possible about the model parameters. Informative priors can be conjugate \citep{diaconis1979conjugate}, mixtures of conjugate priors \citep{dalal1983approximating}, scientifically motivated \citep{tak2018proper, lemoine2019moving}, based on previous data, or in the case of empirical Bayes, based on the data at hand (often called data-driven priors) \citep{carlin2000bayes,maritz2018empirical}. Non-informative priors can be improper or ``flat'', weakly-informative, Jeffrey's priors \citep{tuyl2008comparison}, or other reference distributions. Conjugate priors are sometimes defined to be non-informative. 

One popular choice for a non-informative prior is an \textit{improper} prior --- a prior that is not a probability distribution and in particular does not integrate to one. Good introductions to improper priors are available in the statistics literature \citep{gelman2013bayesian, gelman2017prior}. An example of an improper prior is a flat prior on an unbounded range, e.g., Unif(0, $\infty$) or Unif($- \infty$, $\infty$). 
When an improper prior has been adopted, it is  imperative to check whether the resulting posterior is a proper probability distribution before making any inference. Without posterior propriety the analysis has no probability interpretation. Empirical checks may not be sufficient; posterior samples may not reveal any evidence of posterior impropriety, forming a seemingly reasonable distribution even when the posterior is actually improper \citep{hobert1996propriety, tak2018proper}.

Research on quantifying prior impact is active \citep[e.g., effective prior sample size][]{clarke1996,reimherr2014,jones2020quantifying} as is the discussion on choosing a prior in the context of the likelihood \citep{reimherr2014,gelman2017prior,jones2020quantifying}.

In astronomy, there is a tendency for scientists to adopt non-informative prior distributions, perhaps because informative priors are perceived as too subjective or because there is a lack of easily quantifiable information about the parameters in question. However, all priors provide \textit{some} information about the likely values of the model parameter(s), even a ``flat'' prior. Notably, a flat prior is non-flat  after a transformation. For instance, in our example (Section~\ref{sec:exprior}) a ``non-informative'' uniform prior distribution on the parallax of a star is actually quite informative in terms of distance (third panel, Figure~\ref{fig:prior_to_post}). Thus, we recommend carefully considering what direct or indirect information is available about the value of a parameter before resorting to default or non-informative priors; in astronomy, we usually have at least a little information about the range of allowed or physically reasonable values. Even when a non-informative prior does seem appropriate, checking that the chosen distribution is consistent with known physical constraints is essential. 

Complete descriptions and mathematical forms of prior distributions, including the values of hyperparameters defining these distributions, help promote reproducibility and open science. Unfortunately, a recent meta-analysis of the astronomical literature showed that prior definitions are often incomplete or unstated \citep{tak2018proper}, making it difficult for others to interpret results.

To summarize, good practices in the context of priors are: (1) choosing informative priors when existing knowledge is available, (2) choosing priors with caution if there is no prior knowledge, (3) testing the influence of alternative priors (see the discussion of sensitivity analyses in Section \ref{sec:posterior}), and (4) explicitly specifying the chosen prior distributions for clarity and reproducibility.

\subsubsection{Parallax Example: choosing a prior}\label{sec:exprior}

\begin{figure*}
    \centering
    \includegraphics[width=\textwidth, trim=0 5cm 0 3cm, clip]{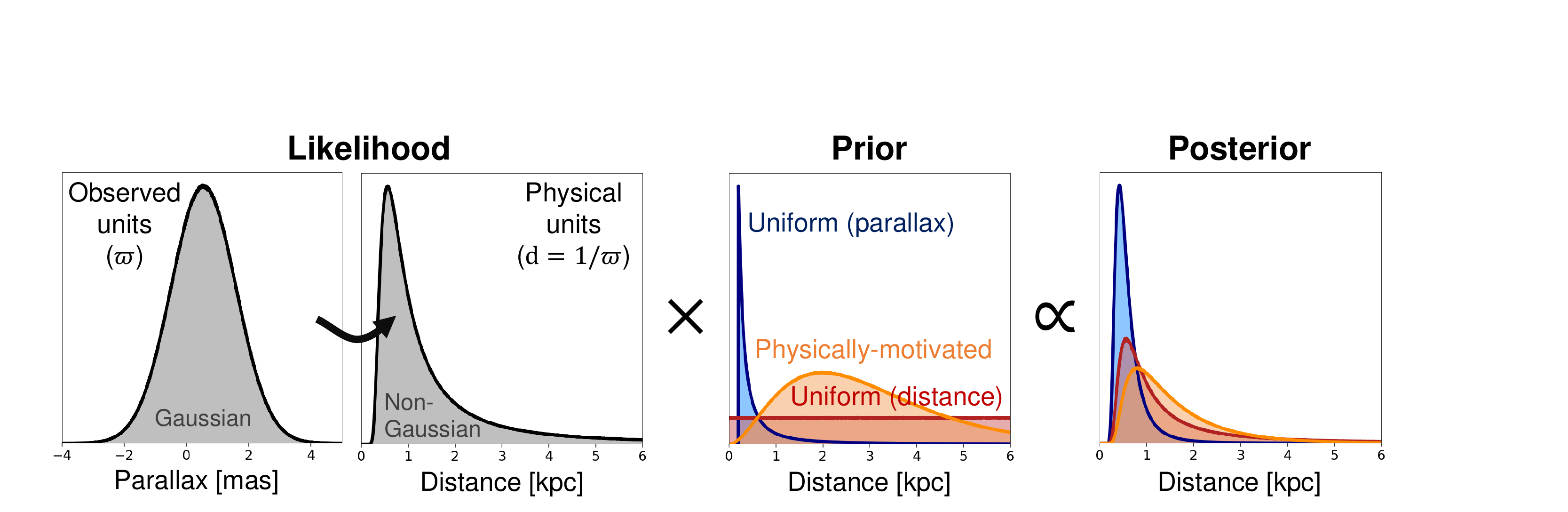}
    \caption{Bayesian inference of the distance $d=1/\varpi$ to a star based on the measured parallax $y$. \textit{Far left:} The likelihood for parallax $\varpi$ is normal with variance assumed known. \textit{Center left:} A transformation of parameters from $\varpi$ to $d$ gives a non-normal PDF. Note that a non-negativity constraint was applied to the distribution of $d$. \textit{Center right:} We highlight three possible priors $p(d)$ over the distance: uniform in parallax $\varpi=1/d$ (blue), uniform in distance $d$ (red), and a physically-motivated prior \citep{BailerJones2018AJ} (orange). \textit{Far right:} The posteriors that correspond to each of the three priors.} 
    \label{fig:prior_to_post}
\end{figure*}

A naive choice of prior on the true parallax $\varpi$ is $p(\varpi) \propto {\rm constant}$, an improper prior that assigns equal density to all values of $\varpi$ from $(0, +\infty)$. A straightforward way to be more informative and proper  is to instead define a truncated uniform prior, where $\varpi$ is uniformly distributed between $\varpi=(\varpi_{\rm min}, \varpi_{\rm max})$ so that
\begin{equation}
    p(\varpi) \propto 
    \begin{cases}
    {\rm constant} & \varpi_{\rm min} < \varpi < \varpi_{\rm max} \\
    0 & {\rm otherwise} 
    \end{cases}, 
    \label{eq:truncuniform}
\end{equation}
or equivalently,
\begin{equation}
    \varpi \sim {\rm Unif}(\varpi_{\rm min}, \varpi_{\rm max})
\end{equation}
Here, $\varpi_{\rm min}$ and $\varpi_{\rm max}$ are hyperparameters set by the scientist (e.g., using some physically-motivated cutoff for $\varpi_{\rm min}$ and the minimum realistic distance to the star for $\varpi_{\rm max}$). Thus the prior in Equation~\ref{eq:truncuniform} can be regarded as weakly-informative because some physical knowledge is reflected in the bounds. Similarly, we could instead define a uniform prior on  distance:% might still
\begin{equation}
    p(d) \propto 
    \begin{cases}
    {\rm constant} & d_{\rm min} < d < d_{\rm max} \\
    0 & {\rm otherwise}
    \end{cases},
    \label{eq:truncuniform2}
\end{equation}
or equivalently,
\begin{equation}
    d \sim {\rm Unif}(d_{\rm min}, d_{\rm max})
\end{equation}
where $d_{\rm min}=1/\varpi_{\rm max}$ and $d_{\rm max}=1/\varpi_{\rm min}$. Like Equation~\ref{eq:truncuniform}, this prior can also be regarded as weakly-informative. However, both display drastically different behavior as a function of $d$ (see Figure~\ref{fig:prior_to_post}), which highlights how the interpretation of non-informative (or weakly-informative) priors may change depending on the choice of parameterization.
%Since we are concerned about the distance parameter instead of 

While the prior in Equation~\ref{eq:truncuniform2} may appear non-informative (in a sense that it is uniform), it actually encodes a strong assumption about the number density of stars $\rho$ as a function of distance. The prior implies that we are just as likely to observe stars at large distances as we are at smaller distances. However, as we look out into space, the area of the solid angle defined by the distance $d$ increases, and this in turn implies that the stellar number density is decreasing with distance. Thus, Equation~\ref{eq:truncuniform2}, which says that all distances are equally likely, implies that there are fewer stars per volume at large distances than stars per volume at small distances. 

\cite{BailerJones2018AJ} introduced a better prior for the parallax inference problem, which we outline briefly and reproduce here. The physical volume ${\rm d}V$ probed by an infinitesimal solid angle on the sky ${\rm d}\Omega$ at a given distance $d$ scales as the size of a shell so that ${\rm d}\Omega \propto d^2 $. This means that, assuming a constant stellar number density $\rho$ everywhere, a prior behaving as $p(d) \propto d^2$ is more appropriate. However, we can go one step further --- we know that our Sun sits in the disk of the Galaxy, and that the actual stellar density $\rho$ as we go radially outward in the disk should decrease as a function of distance. Assuming we are looking outward, and that the stellar density decreases exponentially with a length scale $L$ (so that for a given distance we have $p(\rho|d) \propto e^{-d/L}$) the prior on distance is
\begin{equation}
    p(d) \propto 
    \begin{cases}
    d^2 e^{-d/L} & d_{\rm min} < d < d_{\rm max} \\
    0 & {\rm otherwise} 
    \end{cases},
    \label{eq:EDSD}
\end{equation}
which is the density function of a truncated Gamma($3, L$) distribution. The scientist using Equation~\ref{eq:EDSD} would need to choose and define the three hyperparameters $d_{\rm min}, d_{\rm max},$ and $L$. Equation~\ref{eq:EDSD} is the exponentially decreasing space density prior of previously presented in \cite{BailerJones2018AJ}. Figure~\ref{fig:prior_to_post} illustrates all three priors discussed here.

\subsection{Posterior distributions}\label{sec:posterior}

% analytically or computationally
The posterior distribution of Equation~\ref{eq:Bayes} is the focus of Bayesian inference. Once the prior distribution(s) and the likelihood function are specified, the posterior distribution is uniquely determined. Often, the denominator quantity $p(\bm{y})$ is not available analytically. In this case, $p(\bm{y})$ can be estimated by numerical integration. Samples drawn from $p(\bm{\theta}|\bm{y})$ can be used to estimate properties of the posterior. A popular approach for obtaining samples is to construct a Markov Chain whose stationary distribution is designed to match the target distribution $p(\bm{\theta}|\bm{y})$, which is known as Markov chain Monte Carlo (MCMC). The canonical example is the Metropolis-Hasting algorithm \citep{metropolis1949monte,metropolis1953equation,hastings1970monte,gelman2013bayesian}, but there are many variations, some of which are designed to address specific challenges such as sampling high-dimensional or multi-modal target distributions; see  \cite{brooks2011handbook} for details.

The posterior distribution enables inference of model parameters or of quantities that can be derived from model parameters. For example, the posterior mean $E(\theta\mid y)$ is a point summary for $\theta$. The posterior distribution can also be used to define \emph{credible intervals} for parameters that provide a range of probable values. We stress that credible intervals are not confidence intervals; a 95\% credible interval suggests that there is a 95\% {\it probability} that the parameter lies within the specified range given our prior beliefs, the model, and the data, whereas a 95\% confidence interval suggests that if similar intervals are properly constructed for multiple datasets then 95\% of them are expected to contain the true (fixed) parameter value. 

\begin{figure*}
  \begin{center}
    \includegraphics[width=\textwidth, trim=0 3cm 0 1cm, clip]{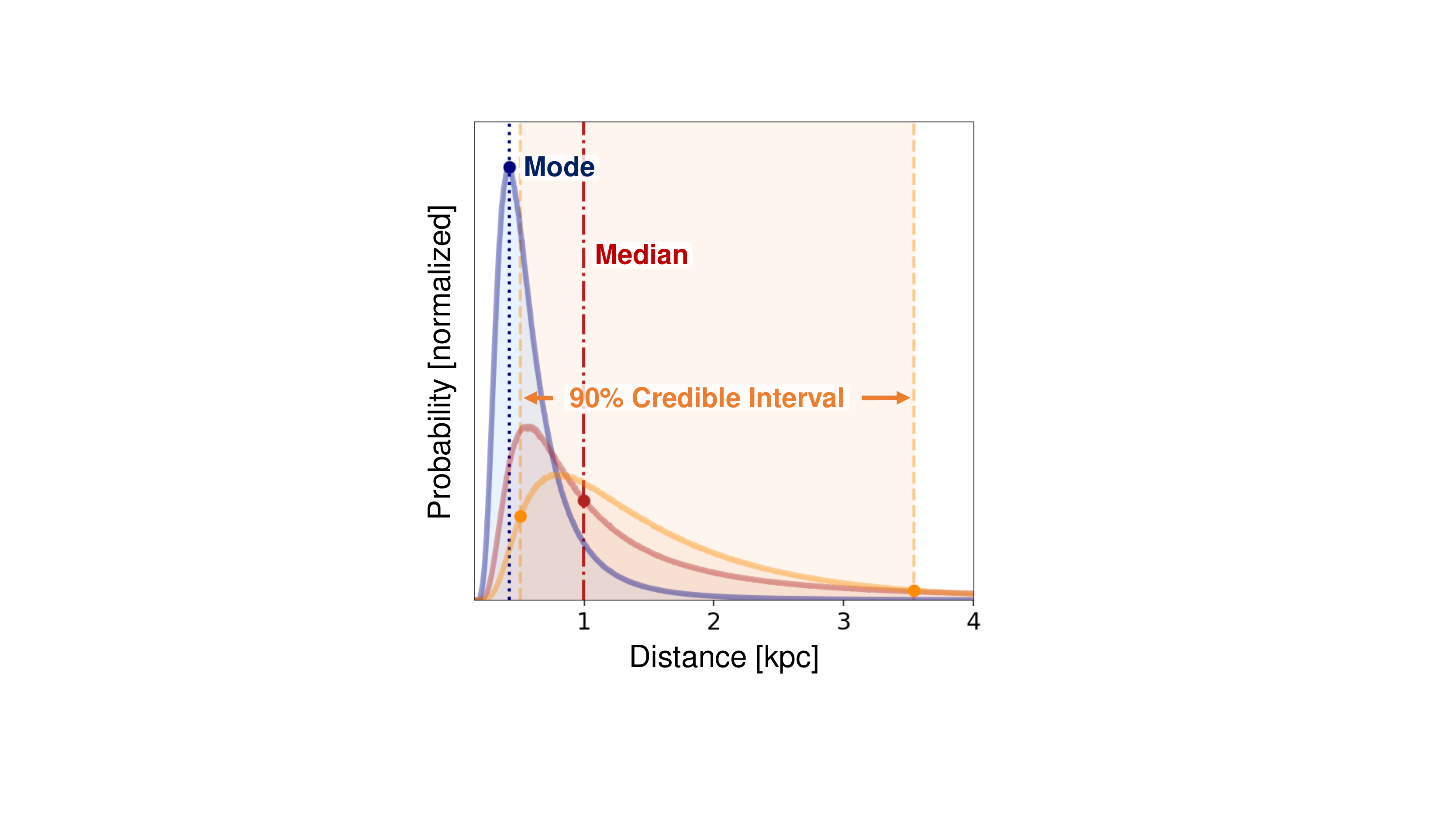}
  \end{center}
  \caption{The three posterior distributions corresponding to each prior distribution shown in Figure~\ref{fig:prior_to_post}: uniform in parallax prior (blue curve), uniform in distance prior (red curve), and physically-motivated prior (orange curve). Also shown are one summary statistic for each posterior:  the mode (blue dotted line), the median (red dotted-dashed line), and the 90\% credible interval (orange dashed lines and shaded region).}\label{fig:post_summary}
\end{figure*}

Depending on the characteristics of the posterior distribution, we emphasize that point summaries and intervals may not provide a complete description of uncertainty (e.g., for multi-modal posteriors). Here, visualizations of the posterior can provide a more comprehensive picture (see Figure~\ref{fig:post_summary}). Recommendations and open source software packages containing visualization tools for Bayesian analysis can be found in the statistics literature \citep{Gabry2019,GabryRpackage, arviz_2019, vehtari2020}. Projections of the joint posterior distribution into two parameter dimensions --- also colloquially referred to as a \textit{corner plot} in astronomy literature --- is the most common visualization tool. Drawing credible regions or contours on these types of visualizations are also helpful, although defaulting to a ``1-sigma'' credible region is not always appropriate (i.e., when the distributions are non-Gaussian).

The posterior distribution also provides a useful way to obtain estimates and credible intervals for other quantities of physical interest. For example, if a model has parameters $\bm{\theta} = (\alpha, \beta, \kappa)$ and there is some physical quantity described by e.g., $\gamma = \beta^2/\alpha e^{\kappa}$, then for every sample of $\bm{\theta}$, a sample of $\gamma$ can be calculated. Thus, a \textit{distribution} of the physically interesting quantity $\gamma$ is obtained, which can also be used to obtain point estimates and credible intervals. In other words, in the Bayesian paradigm, uncertainties in each model parameter are naturally propagated to uncertainties in derived physical quantities in a coherent way.

Posterior distributions can be complicated in shape (e.g., asymmetric, with multiple modes). This can create computational challenges in cases where the posterior cannot be derived in closed form. Fortunately, many algorithms have been developed for approximating the posterior distribution. Different algorithms perform well for different characteristics of the posterior, and therefore prior knowledge of what we might expect the posterior to look like, as well preliminary explorations, are often valuable in practice. In addition to the MCMC sampling algorithms already mentioned, a number of other techniques have been developed. For example,  integrated nested Laplace approximations \citep{rue2009approximate}   approximates posterior distributions,   variational Bayes methods
\citep{jordan1999introduction,blei2003latent,hoffman2013stochastic}, and approximate Bayesian computation \citep{beaumont2009adaptive, marin2012approximate, weyant2013likelihood,akeret2015approximate,ishida2015cosmoabc,
beaumont2019approximate} are possible alternatives when the likelihood functions are too complicated or expensive to be evaluated.

%MCMC-based solutions 
% for dealing with intractable likelihood functions, e.g., 
\subsubsection{Parallax Example: inferring the distance to a star}

In our running example, we are interested in inferring the parameter for the distance $d=1/\varpi$ given the measured parallax $y$ and its associated measurement uncertainty $\sigma$ (which we treat as known). From Bayes' theorem, the posterior is 
\begin{equation}
    p(d | y) \propto p(y | d) p(d).
\end{equation}
For the three priors discussed previously, this corresponds to the following posteriors:
\begin{align}
     {\rm Equation~\ref{eq:truncuniform}}~~~&\Rightarrow~~~ p(d | y) \propto \frac{1}{d^2} \exp\left[{-\frac{(y - 1/d)^2}{2\sigma^2}}\right] \\
    {\rm Equation~\ref{eq:truncuniform2}}~~~&\Rightarrow~~~ p(d | y) \propto \exp\left[{-\frac{(y - 1/d)^2}{2\sigma^2}}\right]\\
    {\rm Equation~\ref{eq:EDSD}}~~~&\Rightarrow~~~ p(d | y) \propto d^2 \exp\left[{-\frac{d}{L}-\frac{(y - 1/d)^2}{2\sigma^2}}\right],
\end{align}
for $d_{\rm min} < d < d_{\rm max}$ (and $0$ otherwise). While none of these have analytic solutions for point estimates or credible intervals, they can be computed using computational techniques. Approximations to these three posterior distributions are show in Figures~\ref{fig:prior_to_post} and \ref{fig:post_summary}.  In this illustration, though each resulting posterior distribution is right-skewed, the shape is notably different for each considered prior distribution.

\subsubsection{Extended Example: inferring the distance to a \textbf{cluster} of stars}\label{sec:extendedexample}

We now extend our example to infer the distance to a cluster of stars, based on the collection of parallax measurements of each individual star. Assuming that there are $n$ stars located at approximately the same distance $d_{\rm cluster}$ and that the measured parallaxes $\bm{y} = \{y_1, y_2, \dots, y_n\}$ to each star are independent given $d_{\rm cluster}$,  our combined likelihood is the product of the individual likelihoods
\begin{equation}
p(y_1,y_2, \dots, y_n | 1/d_{\rm cluster}) = \prod_{i=1}^{n} p(y_i | 1/d_{\rm cluster}),
\end{equation}
where the individual likelihoods are defined following Equation \ref{eq:model}. We assume that the measurement uncertainties $\sigma_i$ are  known constants.
Our posterior is 
\begin{equation}
    p(d_{\rm cluster} | \bm{y}) \propto p(d_{\rm cluster}) \prod_{i=1}^{n} p(y_i | 1/d_{\rm cluster}).
\end{equation}

The product of $n$ independent Gaussian densities with known variances is a Gaussian density with precision parameter $\tau^{-2} = \sum_{i=1}^{n} \sigma_i^{-2}$ and mean parameter
$1/d_{\rm cluster}$.  The observed parallaxes can be combined to obtain an effective parallax $y_{\rm eff} = \tau^2\sum^{n}_{i=1}y_i/\sigma^2_i$, and thus,
\begin{equation}
    p(d_{\rm cluster} | \bm{y} ) \propto p(d_{\rm cluster}) \exp\left[{-\frac{(y_{\rm eff} - 1/d_{\rm cluster})^2}{2\tau^2}}\right].
\end{equation}

\begin{figure*}
    \centering
    \includegraphics[width=\textwidth, trim=0 1cm 0 1cm, clip]{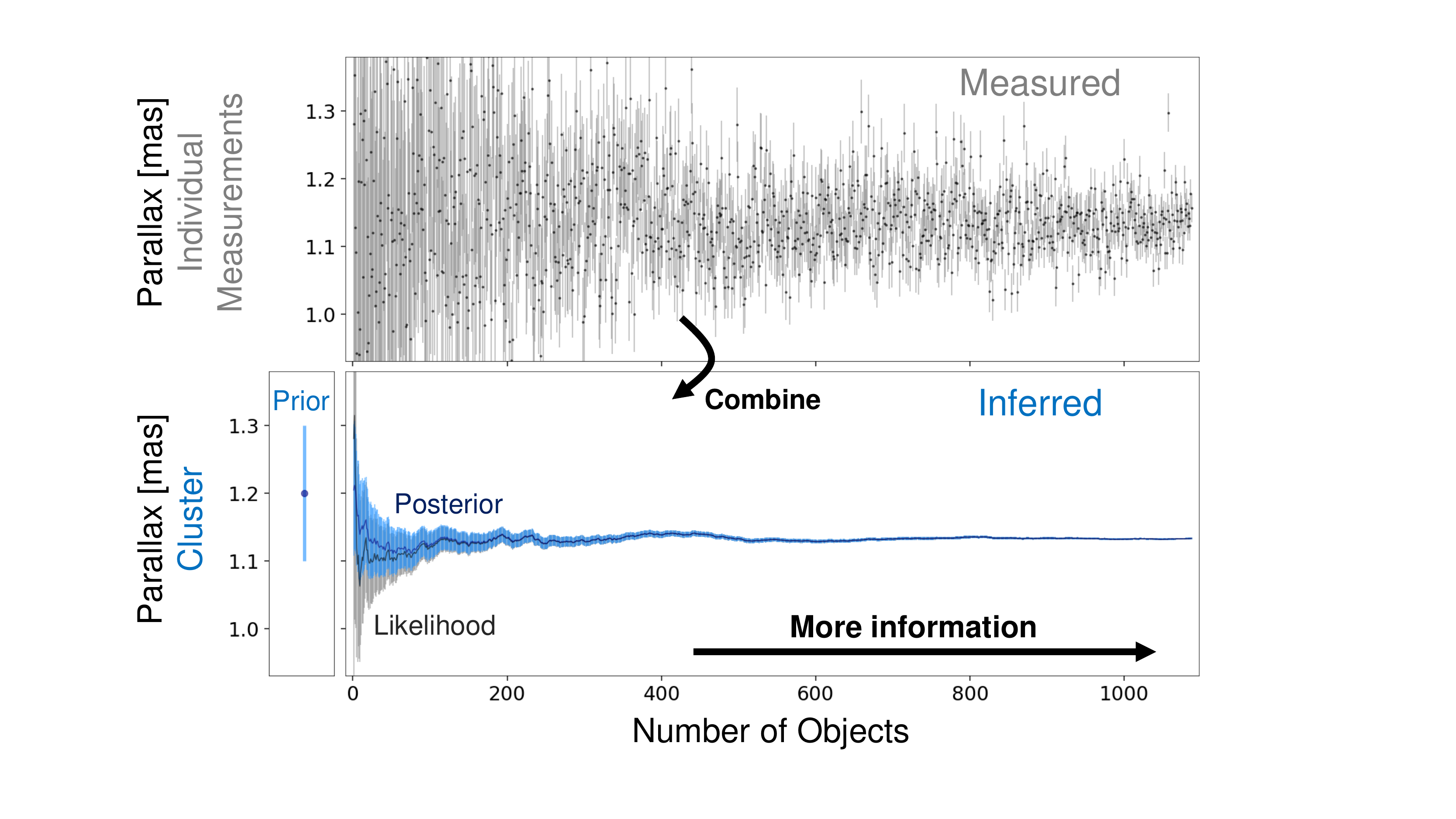}
    \caption{\textbf{Extended Example for Open Cluster M67.} An extension of the example shown in Figure~\ref{fig:prior_to_post} illustrating how to infer the distance to an open cluster (M67) based on parallax measurements of many stars. \textit{Top:} Parallax measurements (gray) for likely cluster members (based on proper motions), sorted by their observed signal-to-nose ratio $\varpi_{\rm obs}/\sigma_\varpi$. \textit{Bottom:} The joint likelihood (gray) and posterior (blue) for the cluster parallax $\varpi_{\rm cluster} = 1/d_{\rm cluster}$ as more and more stars are added to our analysis. The (Gaussian) prior distribution on the cluster's parallax is illustrated in the narrow left panel. When there is only a small number of stars, the location of the prior has a substantial impact on the posterior. However, as more stars are added, the information from the data dominates.}
    \label{fig:prior_to_post2}
\end{figure*}

The estimated posterior distribution over $\varpi_{\rm cluster} = 1/d_{\rm cluster}$ for a nearby cluster of stars (M67) using data from \textit{Gaia} DR2, and using a conjugate Gaussian prior for $\varpi_{\rm cluster}$, is shown in Figure \ref{fig:prior_to_post2}. The top panel of Figure~\ref{fig:prior_to_post2} shows the individual parallax measurements of stars in M67, sorted by their signal-to-noise values. The bottom panel shows the assumed prior distribution (narrow left panel) and the (estimated) posterior distribution for the cluster's parallax, as more and better data are added to the analysis.

\textbf{Note:} A prior over $p(d_{\rm cluster})$, which governs the distribution of clusters of stars, is not the same as a prior over $p(d)$, which governs the distribution of individual stars. While it might be reasonable to assume these are similar, they are not interchangeable quantities and may indeed follow different distributions. Realizing the differences in priors between various scenarios such as these is key to building good models and subsequently making good inferences.

\subsection{Posterior Predictive Checking}\label{sec:pcheck}

After obtaining the posterior distribution, it is  recommended to assess  the adequacy of the model using \emph{posterior predictive checks} \citep{gelman1996ppc}, which compare the empirical distribution of the data to the distribution described by the Bayesian model. The posterior predictive distribution is the posterior distribution of hypothetical future data ($\bm{\Tilde{y}}$) under the chosen model and given the previously collected data:
\begin{equation}
    p(\bm{\Tilde{y}}|\bm{y}) = \int p(\bm{\Tilde{y}}, \bm{\theta}|\bm{y})d\bm{\theta}.
    \label{eq:PPD}
\end{equation}
We find that posterior predictive checking is underused in astronomy but can be very useful. Posterior predictive checks not only assess the the adequacy of the model but also simultaneously check any approximations to the posterior. They are a valuable tool for diagnosing issues with computational sampling methods.

In most cases, the posterior predictive distribution is not available in closed form. However, it is possible to generate simulated observations from the posterior predictive distribution and compare these to the original data. For example, for each of the  posterior samples of $\bm{\theta}$, draw a random sample of $\bm{\Tilde{y}}$ and compare these samples to the real data. Significant or systematic differences between the distributions of the real and simulated data may suggest a problem with the model. 

\begin{figure}
\centering
  %  trim={<left> <lower> <right> <upper>}
    \includegraphics[width=0.75\textwidth, trim=10cm 1cm 5cm 1cm, clip]{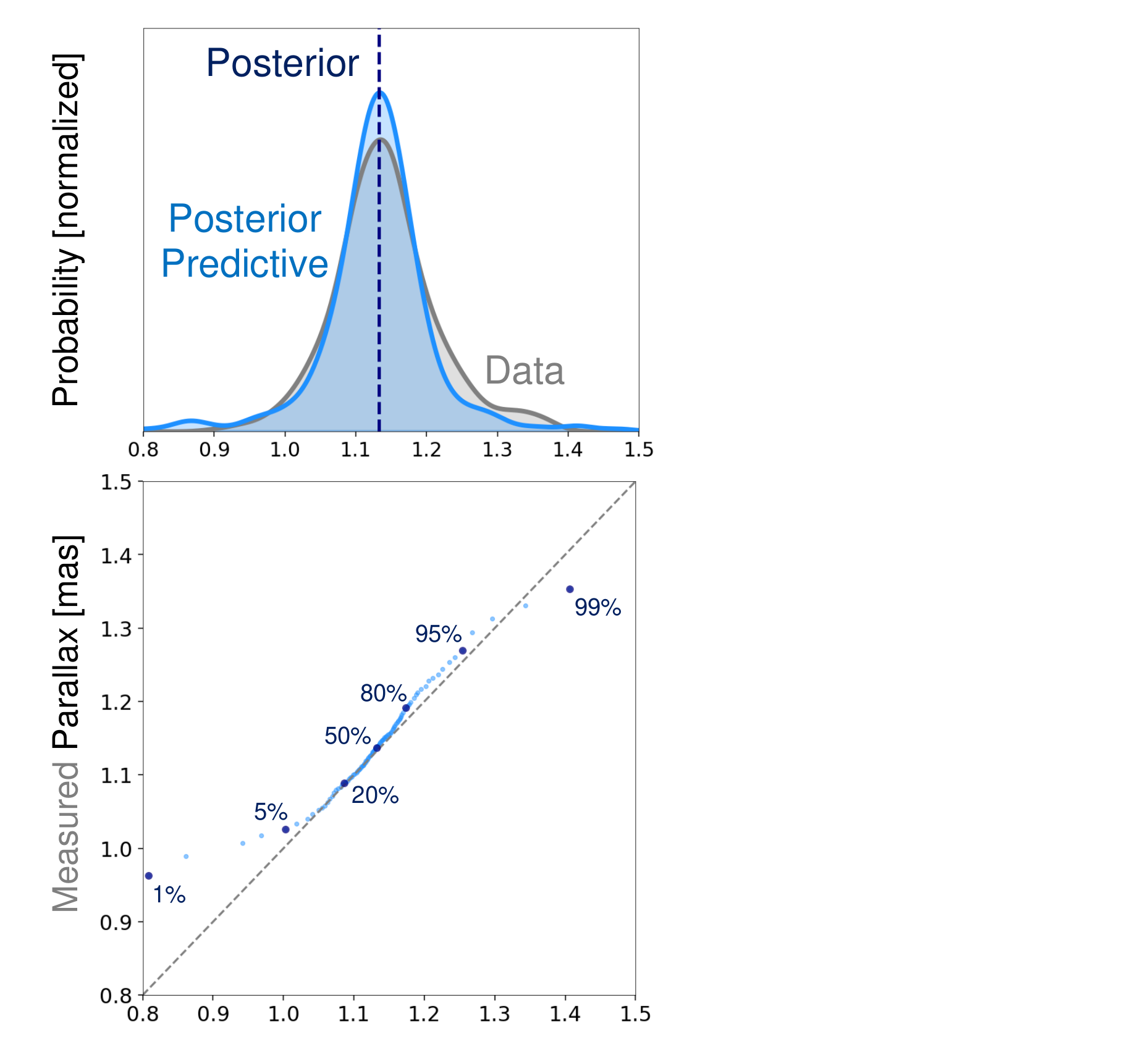}
  \caption{\textbf{Quantile-Quantile (\textit{Q-Q}) Plot.} This figure demonstrates a way to perform posterior predictive checking for the model shown in Figure~\ref{fig:prior_to_post2}. \textit{Top:} The distribution of parallax measurements from the data (gray) and simulated values from the posterior predictive (light blue). The posterior mean is indicated using the dashed dark blue line. The distributions appear relatively consistent with each other by eye, but a quantile-quantile (\textit{Q-Q}) plot is more informative and suggests otherwise. \textit{Bottom:} The \textit{Q-Q} plot of the quantiles from the posterior predictive simulated parallax data ($x$-axis) and of the observed parallaxes ($y$-axis). If the real and simulated data followed the same distribution, then the quantiles would lie on the one-to-one line. However, strong discrepancies are apparent below $\sim10$th percentile and above $\sim70$th percentile.}
  \label{fig:post_pred}

\end{figure}

It is good practice to perform quantitative and/or graphical comparison between the simulated data and the real data. For graphical comparison, an overlaid density plot could be used (top panel of Figure~\ref{fig:post_pred}), but in general this is a poor choice because it is difficult to judge differences between the overlaid densities visually. For graphical comparison, we recommend instead using a \textit{quantile-quantile }(\textit{Q-Q}) \textit{plot} to characterize any differences (bottom panel of Figure~\ref{fig:post_pred}). 

To construct a \textit{Q-Q} plot, it suffices to compute the quantiles from the original data and from the posterior predictive distribution (or from data simulated from the posterior predictive distribution), and to plot the pairs one against the other (bottom panel, Figure~\ref{fig:post_pred}). If the empirical distribution and the posterior predictive distribution match, then their quantiles should lie along a 1:1 line. Functions to display \textit{Q-Q} plots are common in statistical computing software languages. 

In our parallax example, parallax values in the tails of the simulated and real data distributions show some disagreement (Figure~\ref{fig:post_pred}). The \textit{Q-Q} plot shows this more explicitly that the density plot, as the quantiles do not follow the 1:1 line in the tails of the distribution (below $\sim10$th percentile and above $\sim70$th percentile). Differences in either end of a \textit{Q-Q} plot can be due to chance, but strong deviations from the 1:1 line are usually worth investigating.

There are also other valuable approaches for checking a Bayesian model and the quality of approximations to the posterior distribution. For example, one may set a portion of the data aside, or obtain additional data, and then compare the resulting inference to that from the original data. One may also use multiple methods for approximating the posterior distribution, and compare results. This can help diagnose situations where one or more sampling algorithms did not explore the full parameter space, and consequently fail to include high-probability regions in the posterior samples.

In addition to model and posterior checking, it is important to consider the influence of the prior distribution(s). The rightmost panel of Figure~\ref{fig:prior_to_post} shows three posteriors: each one used one of the three priors discussed in Section~\ref{sec:exprior}. While the posteriors are vaguely similar in shape (e.g., right-skewed), the inferred summary statistics can be quite different (Figure~~\ref{fig:post_summary}).

More generally, investigating how the analysis compares for several different prior distributions is an important technique, often referred to as a \textit{sensitivity analysis}. A sensitivity analysis directly assesses the impact of the prior distribution on the posterior, and for this reason we recommend them --- particularly when the information available to construct a prior is limited. On the other hand, sensitivity analyses can be somewhat ad hoc (e.g., which priors are tried, how they are compared) making it difficult to summarize and compare the prior impact across multiple analyses, instruments, and models. More principled approaches may therefore be preferred or complementary in some scenarios. One such method is to quantify the effective prior sample size (EPSS), i.e., the number of data points that the information provided by the prior distribution corresponds to. 

The EPSS is simple to compute  for conjugate models. For example, if we have the data $y_i\overset{iid}{\sim} N(\mu,\sigma^2)$, for $i=1,\dots,n$, and the conjugate prior distribution $\mu \sim N(\mu_0,\sigma^2/m)$, with known $\sigma^2$, then the posterior distribution of $\mu$ has variance $\sigma^2/(n+m)$. Thus, the effect of the prior is equivalent to that of $m$ samples, and we say that the EPSS is $m$. The statistics literature includes proposals of several methods for extending this idea beyond conjugate models \citep{clarke1996,Morita:2008,reimherr2014,jones2020quantifying} and how to additionally account for location discrepancies, e.g., the value of $|\bar{y}-\mu_0|$ in the preceding example. \cite{clarke1996} and \cite{Morita:2008} use EPSS to quantify the information in the prior in isolation from the data, while \cite{reimherr2014} and \cite{jones2020quantifying} concentrate on the impact of the prior on the specific analysis performed. The latter is typically more relevant in science and more closely coincides with sensitivity analyses.

Good practices outlined in this section can be summarized as (1) using multiple ways to summarize the posterior inference, (2) quantitatively and graphically checking the posterior distribution (e.g., using posterior predictive checks, \textit{Q-Q} plots), and (3) providing evidence that diagnostic checks were completed.

\subsubsection{Extended Example: posterior predictive checks}
We investigate the validity of our model for the distance to a cluster of stars by computing the posterior predictive distribution for the observed stellar parallaxes. While in this case the posterior predictive can be written in closed form (since it is a Gaussian distribution), we also approximate it by simulating values of $d_{\rm cluster}$ from the posterior and then subsequently simulating values for the predicted parallax measurements $\varpi_{{\rm pred}, i}$ given $d_{\rm cluster}$. 

In Figure~\ref{fig:post_pred}, we compare both the distribution and quantiles estimated for the simulated dataset and the observed dataset via a density and \textit{Q-Q} plot respectively. While there are differences, especially in the tails of the distribution, overall the cluster model reproduces most of the observed properties of the data. It would be worth investigating whether these differences persist under different models -- for example, a model in which the distance to each star is not assumed to be identical, or a model in which measurement uncertainty is not assumed to be known exactly.

\subsection{Conclusion}
% "The Conclusions section should briefly summarize the main points of the article and comment on the implications of the most recent work, on open challenges and on future research directions."
We hope that this article has identified, clarified, and illuminated fundamental Bayesian inference notation and techniques from the statistics literature, and in particular, has made a case for fully specifying the model, posterior predictive checking, and the use of underused aids such as the \textit{Q-Q} plot. In summary, we highlight sound practices for conducting Bayesian inference in astronomy as follows: % following recommendations for :
\begin{itemize}
    \item Be explicit about notation, and use appropriate terminology for the interpretation of concepts such as the likelihood and credible intervals, which will help interdisciplinary collaboration and reproducibility. 
    \item Describe the likelihood as a function of the parameters, given the data. 
    \item Use informative priors whenever possible and justified. Carefully consider what direct or indirect information is available about the parameters. 
    \item Use non-informative priors carefully, and assess their properties under parameter transformations.
    \item Test the sensitivity of the posterior distribution to different prior distributions.
    \item Fully specify the Bayesian model in terms of the likelihood, prior, and posterior, and provide open-source code whenever possible.
    \item Perform posterior predictive checks of the model, using visualizations such as \textit{Q-Q} plots where appropriate.
    \item Strive to include all non-negligible contributors to the measurement process.
    
\end{itemize}

We hope that there is a continued growth of interdisciplinary collaborations between astronomers and statisticians in the future. Data from cutting-edge telescopes such as the Vera Rubin Observatory, the James Webb Space Telescope, and many others, have the potential to drive the field of astronomy, but this new information is best understood in the context of existing knowledge and careful statistical inference. Bayesian inference provides a framework in which this type of analysis and discovery can occur. Areas of astronomy where prior information and non-Gaussian based likelihoods are common can especially benefit from Bayesian methods, for example X-ray and gamma-ray astronomy. 

Bayesian inference is a broad topic, and many subtopics were not covered in this article. Ultimately, we hope that this article not only serves as a useful resource, but will also be the inception for a series of more specific papers on Bayesian methods and techniques in astronomy and physics.

\section*{Acknowledgements}

GME acknowledges the support of a Discovery Grant from the Natural Sciences and Engineering Research Council of Canada (NSERC, RGPIN-2020-04554). JCK gratefully acknowledges support from NSF under
Grant Numbers AST 2009528 and DMS 2038556.
DH is supported by the Women In Science Excel (WISE) programme of the Netherlands Organisation for Scientific Research (NWO).

%The Acknowledgements section is not numbered. Here you can thank helpful colleagues, acknowledge funding agencies, telescopes and facilities used etc. Try to keep it short.

%%%%%%%%%%%%%%%%%%%%%%%%%%%%%%%%%%%%%%%%%%%%%%%%%%
\section*{Data Availability}

% The inclusion of a Data Availability Statement is a requirement for articles published in MNRAS. Data Availability Statements provide a standardised format for readers to understand the availability of data underlying the research results described in the article. The statement may refer to original data generated in the course of the study or to third-party data analysed in the article. The statement should describe and provide means of access, where possible, by linking to the data or providing the required accession numbers for the relevant databases or DOIs.

Data used in the running example is provided with permission and courtesy of Phill Cargile (Center for Astrophysics $|$ Harvard \& Smithsonian). 

%\section*{Author contributions statement}
%G.M.E. conceived the paper and led the organization, chaired co-author meetings, structured and reorganized writing contributions, and did the majority of editing. G.E. wrote the Introduction and Conclusion, and also wrote parts of Sections~\ref{sec:notation}-\ref{sec:posterior} inclusive. J.S.S. led the analysis for the running example, writing all code and making all figures, and contributing text to Sections~~\ref{sec:likelihood}, \ref{sec:prior}, and \ref{sec:posterior}. J.C.K., D.F.M., D.H., D.J., A.S., and H.T. contributed substantial writing to Sections~\ref{sec:notation}-\ref{sec:posterior} inclusive. All co-authors contributed equally to discussions about the concepts that would go into the paper, researching appropriate references, and making suggestions for figures and text changes. All authors reviewed the manuscript.


\bibliographystyle{mnras}
\bibliography{refs}

%\begin{thebibliography}{99}
%\end{thebibliography}

%%%%%%%%%%%%%%%%%%%%%%%%%%%%%%%%%%%%%%%%%%%%%%%%%%

%%%%%%%%%%%%%%%%% APPENDICES %%%%%%%%%%%%%%%%%%%%%

% \appendix

% \section{Some extra material}

% If you want to present additional material which would interrupt the flow of the main paper,
% it can be placed in an Appendix which appears after the list of references.

%%%%%%%%%%%%%%%%%%%%%%%%%%%%%%%%%%%%%%%%%%%%%%%%%%

% Don't change these lines
\bsp	% typesetting comment
\label{lastpage}
\end{document}